\documentclass[cameraready]{Interspeech}
\usepackage{booktabs}
\usepackage{amsmath,graphicx}
\usepackage{xcolor}
\usepackage{amsmath}
\usepackage{amssymb}
\usepackage[caption = false]{subfig}
\usepackage{cite}
\usepackage{comment}

\newcommand{\Am}[1]{\mathbf{A}_{#1}}
\newcommand{\Bm}[1]{\mathbf{B}_{#1}}
\newcommand{\Cm}[1]{\mathbf{C}_{#1}}

\title{DAVSS: Distilled Audio-Visual State Space Models}
\author{Saurabhchand Bhati$^*$, Mrudula Athi$^ \dagger$, Amit S. Chhetri$^ \dagger$, James Glass$^*$}

\address{
$^1$ CSAIL, MIT \\
$^2$ Amazon}

\email{\small{sbhati@mit.edu}}

\keywords{State-space models, Audio-visual modeling, Knowledge distillation}

\begin{document}

\maketitle

\begin{abstract}
State-space models (SSMs) distilled from transformer teachers combine the performance of transformers with the efficiency of SSMs. We extend the Transformer-SSM knowledge distillation to a multimodal setting and propose the Distilled Audio-visual State-Space (DAVSS) model. The DAVSS model, 14M parameters, is 12 times smaller compared to transformer-based models such as CAV-MAE, and still outperforms them. 

DAVSS improves over the existing audio-visual models by:
1) Finer input resolution: using smaller patch sizes process the input, compensating for the smaller model size by increasing input sequence lengths. This is supported by the observation that a larger patch size results in lower performance. 
2) Deeper joint modeling: utilizing a larger portion of the model (30\%) for joint audio-visual processing, compared to $<$5\% in CAV-MAE, enabling deeper cross-modal interaction without significantly increasing the computational cost associated with the concatenated audio-visual tokens.

\end{abstract}

\section{Introduction}
Audio and visual modalities share an underlying correspondence despite their differences, and play an important role in our perception of the world.  
Jointly learning from the two modalities has been an active research direction in the multimodal research community~\cite{aytar2016soundnet,arandjelovic2017look,afouras2020self,morgado2021audio,gongcontrastive,haliassos2022jointly,huang2023mavil,sarkar2023self,kim2024equiav,georgescu2023audiovisual,guo2024crossmae,xing2024locality,lin2024siamese,araujo2025cav}.  Transformers have become the dominant modeling paradigm for audio and audio-visual modeling; however, they suffer from quadratic time self-attention operation~\cite{gu2021efficiently,gu2023mamba}. In a multimodal setting such as audio-visual, the tokens from the two modalities are often concatenated, resulting in longer sequence lengths, which is not ideal due to the quadratic nature of attention. Thus, the fusion of audio and visual modalities often remains limited to the last layers~\cite{gongcontrastive,araujo2025cav}.  

\begin{figure}[th!]
    \subfloat[Proposed knowledge distillation framework]{
    \hspace{-1pt}
    \includegraphics[width=0.8\columnwidth]{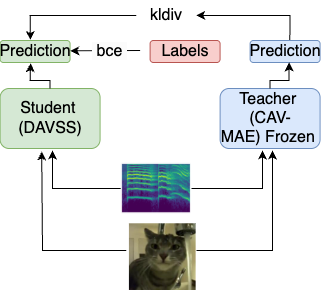}
    }\\
    \centering
    \subfloat[Relative inference speed vs mAP]{
    \includegraphics[width=0.8\columnwidth]
    {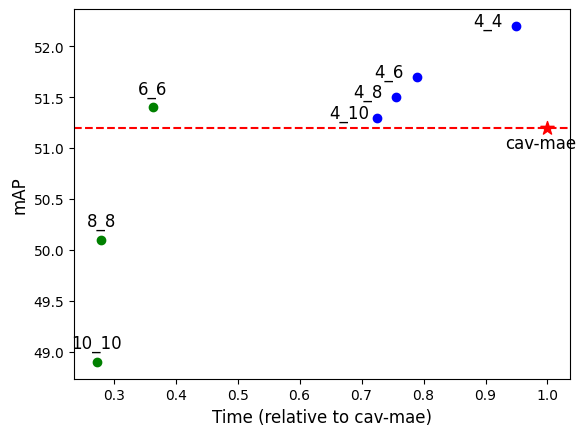}}
    \caption{(a) We apply knowledge distillation in audio-visual SSM training using CAV-MAE as the teacher. (b) Performance of DAVSS models compared to the CAV-MAE model. To measure the inference time, we use 5000 audio-visual pairs. As the number of tokens decreases, the inference speed increases while performance goes down. The model with both audio and visual patch size 6 is significantly faster than the CAV-MAE model while still outperforming it.}
    \label{fig:DASS_overview}
\end{figure}

State-space models have emerged as an alternative to transformers due to their high computational efficiency, especially for long inputs~\cite{gu2021efficiently,gu2023mamba,dao2024transformers,zhu2024vision,liu2024vmamba,bhati2024dass,hamza2024audio,lin2024audio,zhang2025mamba,jiang2025speech}. However, their performance on the classification task, such as audio classification, lagged behind the transformer-based models~\cite{hamza2024audio,lin2024audio}. Recent approaches use knowledge distillation from transformer-based teachers to state-space models to leverage the existing transformer-based model to boost the performance of SSMs~\cite{bhati2024dass}. 
Distilled audio state-space (DASS) models combine the strengths of the two architectures~\cite{bhati2024dass}. DASS achieves high performance like the transformer models and is efficient like the state-space models. DASS achieves state-of-the-art (SOTA) performance on the audio-classification task and outperforms the existing transformer and state-space-based approaches. 
In this work, we extend the knowledge distillation framework to a multimodal, i.e., audio-visual setting. 
We show that the DAVSS model is significantly smaller (specifically 3.5x smaller) compared to the CAV-MAE model and still outperforms the CAV-MAE model on the AudioSet audio-classification task~\cite{gongcontrastive}.

Features combined from audio-visual modalities result in longer sequences, and to reduce the computational cost, a smaller portion of the audio-visual models process the data jointly. Thus, these models mainly focus on modelling these two modalities independently. With state-space based DAVSS, we can handle longer sequences and use a larger portion of the model for modelling the audio-visual data jointly. In CAV-MAE, less than 5\% of the total parameters are in the joint layers, whereas DAVSS contains 30\% of the total parameters in the joint layers.

Despite using a longer input sequence, DAVSS is still faster than CAV-MAE due to the efficient nature of SSMs. To further increase the speed, we vary the resolution of the DAVSS models, which allows us to control the throughput of the models. Increasing the patch size allows us to have models that work twice as fast as the CAV-MAE model while still outperforming the CAV-MAE model. As the size of the audio spectrogram is larger than the visual input, the audio input produces more tokens for the same patch size. As observed in Figure 1 (b), the patch size for the audio branch affects the model inference speech more than the patch size for the visual branch.

\section{DAVSS: Distilled Audio-Visual State-space models}

\subsection{State-Space Models}
Structured state space sequence models (S4)~\cite{gu2021efficiently} are inspired by classical state-space models and are broadly related to recurrent neural networks (RNNs) and convolutional neural networks (CNNs).
The continuous state-space models map a 1-D input sequence $ x(t) \in \mathbb{R} \rightarrow y(t) \in \mathbb{R}$ through a hidden state $h(t) \in \mathbb{R}^{N}$ via the following linear ordinary differential equations:
\begin{align}
    & \mathbf{h'(t)} = \Am{}\mathbf{h(t)} + \Bm{}x(t), \\
    & y(t) = \Cm{}\mathbf{h(t)}   
\end{align}
where $ \mathbf{A} \in \mathbb{R}^{N\times N}$ is called the evolution parameter and $ \mathbf{B} \in \mathbb{R}^{N\times 1}, \mathbf{C} \in \mathbb{R}^{1\times N}$ are called the projection parameters. 

To model SSMs with neural networks, a discretization method is applied. Commonly, zero-order hold (ZOH) is used for discretization, which uses a timescale parameter $\Delta$ to transform the continuous parameters $\Am{}$, $\Bm{}$ to discrete parameters $\overline{\Am{}}, \overline{\Bm{}}$ as follows:
\begin{align}
    &\bar {\Am{}} = \exp(\Delta \Am{}), \\
    &\bar {\Bm{}} = (\Delta \Bm{})^{-1}(\exp(\Delta \Am{}) - \mathbf{I})\Delta \Bm{}
\end{align}

After the discretization step, the state-space equations can be rewritten as:
\begin{align}
    & h_{t} = \overline{\Am{}}h_{t-1} + \overline{\Bm{}} x_{t} \\
    & y_{t} = \Cm{} h_{t}
\end{align}

One major advantage of SSMs is that we can view them as either CNNs or RNNs depending on the task~\cite{gu2021efficiently}.
While SSMs showed remarkable performance on language modeling tasks, the linear time-invariant nature of SSMs limited their ability to capture contextual information well and performed poorly on content-based reasoning tasks~\cite{gu2023mamba}. 

\begin{figure*}[t!]
    \centering
    \includegraphics[width=0.80\linewidth]{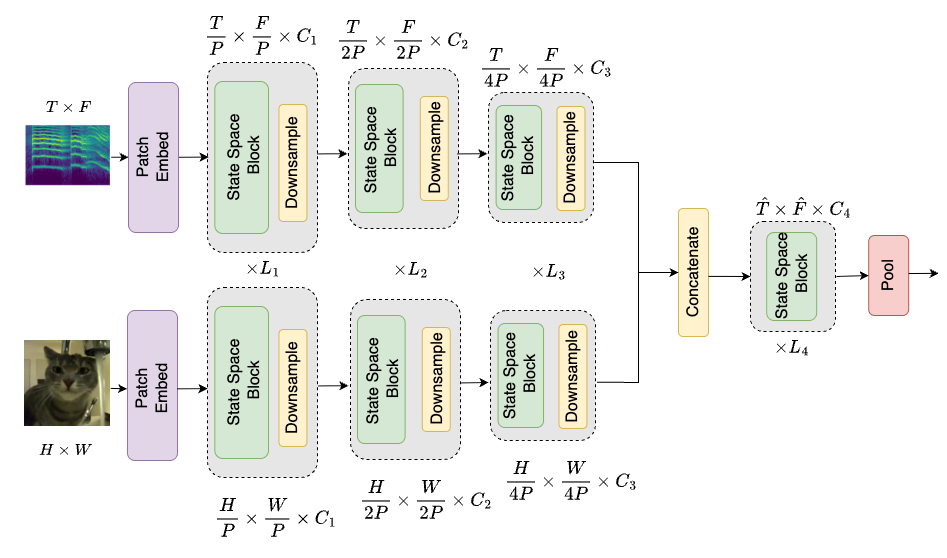}
    \caption{Overview of the audio and visual encoder and the joint modelling blocks in DAVSS}
    \label{fig:DAVSS_arch}
    \vspace{-2mm}
\end{figure*}

To tackle the limitations of SSMs and improve the performance on the contextual reasoning tasks, Gu et al.~\cite{gu2023mamba} proposed a parameterization method to make SSM parameters input-dependent and proposed selective scan S4. However, the state parameters are now input dependent, which means the convolutional view can not be used. The recurrent view is still applicable, but it can not be trained in parallel like the convolutional view. Hardware-aware optimizations and parallel scan algorithms are used to speed up training and inference speeds. 

\subsection{DAVSS}

The overview of our proposed DAVSS model is shown in Figure~\ref{fig:DASS_overview} while Figure~\ref{fig:DAVSS_arch} shows the detailed architecture of the state-space student model. 

The model contains two modality-specific encoders for audio and visual modality, which process the two modalities independently. The features from the two modalities are merged and processed by a joint block. The two modality-specific encoders are identical, and each encoder can be subdivided into three stages. Each stage consists of a state-space block and a patch merging-based downsampling layer. The model progressively reduces the spatial dimensions and increases the number of features. In the end, we combine the features generated by the two encoders, and a group processes the joint features. A pooling method generates the final embedding for the audio-visual input, which is then passed into a classifier to generate the final output of the model. 

Specifically, the visual encoder takes a two-dimensional image $ X \in \mathbb{R}^{H \times W} $ as input, then a patch embedding layer extracts two-dimensional feature patches with spatial dimensions $\frac{H}{P} \times \frac{W}{P} \times C_1$. The first group processes the features at this scale and downsamples the features to $\frac{H}{2P} \times \frac{W}{2P} \times C_2$ dimensions. This continues for another two blocks, and the visual encoder generates features with a spatial dimension $\frac{H}{4P} \times \frac{W}{4P} \times C_3$. Similarly, the audio encoder takes a two-dimensional spectrogram $ X \in \mathbb{R}^{T \times F} $ as input and generates features with spatial dimension $\frac{T}{4P} \times \frac{F}{4P} \times C_3$. To combine the 2 two-dimensional features generated by the audio and visual branches, we first reshape them to one dimension, concatenate the features, and then reshape them back to two-dimensional features. For example, for a patch size of 4, the audio and the visual encoder generate 32x4 and 7x7 dimensional features, and these are then combined into a 59x3 feature block for the last group. We use a pooling method to summarize information into a single embedding, which is then passed into a linear layer to generate the final classification output.  

We utilize the existing transformer-based models to train and boost the performance of SSM. We use knowledge distillation from a transformer-based teacher model (CAV-MAE) to distill knowledge into an SSM-based student (DAVSS). We pass the same audio-visual input to the student, DAVSS, and the teacher, the CAV-MAE model, and generate the outputs from the two models. The student model is trained to predict the ground truth labels and match the output of the teacher model. The overall loss for the model is $
    \mathcal{L} = 0.5(\mathcal{L}_{bce}(y,\hat{y}_{stu}) + \mathcal{L}_{kldiv}(\hat{y}_{teach},\hat{y}_{stu}))$,
where $\mathcal{L}_{bce}$ is the binary cross-entropy loss between the output of the student DAVSS model $\hat{y}_{stu}$ and the ground truth labels $y$, and $\mathcal{L}_{kldiv}$ is the KL-divergence between $\hat{y}_{stu}$ and the output of the teacher model $\hat{y}_{teach}$.

\section{Experiments}

We use the audio-visual based audio-event classification task for evaluating our proposed DAVSS model. For this, we use the AudioSet dataset to train and evaluate our models. The dataset and the training details are described next.

\subsection{Dataset and Training Details}
AudioSet~\cite{gemmeke2017audio} contains approximately 2 million 10-second videos extracted from YouTube videos. These videos are labeled from a set of 527 labels. The full training (AS-2M) and the evaluation datasets contain 2M and 22k data points, respectively. VGGSound~\cite{chen2020vggsound} contains approximately 200K 10-second video clips with 309 sound events. The train and test subset contains 183K and 15K data points.

For audio and visual inputs, we follow the preprocessing steps from the CAV-MAE model~\cite{gongcontrastive}. Specifically, we use 10-second videos along with the parallel audio. The audio waveform is first converted to a 1024($T$) x 128($F$) spectrogram. We sample the video at 1 frame per second (FPS) and generate 10 RGB frames. During training, a frame is randomly selected as input to the model. During inference, we average the model prediction of each RGB frame.
For each RGB frame, we resize and center crop it to 224($H$) × 224($W$).

Following DASS~\cite{bhati2024dass}, we increase the learning rate to 1e-4. We train the models for 10 epochs with a learning rate scheduler, which lowers the learning rate by half every epoch starting from the second epoch. We use a batch size of 36 and a mixup probability of 0.5 to augment the data to train the models. We experiment with three model sizes: DAVSS-Tiny, DAVSS-Small and DAVSS-Medium, containing (2, 2, 8), (2, 2, 8) and (2, 2, 15) layers in the three groups in the two modality-specific encoders, respectively. The models have feature dimension of $C_1,C_2,C_3,C_4$ = (64,128,256,512), (96,192,384,768) and (96,192,384,768) across the four groups. DAVSS-Tiny, DAVSS-Small and DAVSS-Medium contain 14.4, 46.9 and 84.4 million total parameters, respectively.

\subsection{Modality specific depth}
Typically, the fusion of audio and visual features is done at the very end of the models. For example, the CAV-MAE model uses 11 out of 12 layers for the modality-specific encoders and just the last layer for joint modeling. This minimizes the number of layers processing the joint data with longer sequence lengths, as transformers scale quadratically with the sequence lengths. SSMs can model longer sequences efficiently, which means we can utilize more parameters for joint modeling. 

In this section, we explore the impact of various ratios of joint vs modality-specific layers. Since fusion at different layers will change the number of layers and therefore the parameters in the model, which impacts the performance. We modify the model so that both the audio and visual encoders have four groups, and we vary the location where features are fused. This ensures that the total number of parameters is the same across the different experiments. For example, if the features are being merged after the third group, then the fourth group in both audio and visual encoders processes the joint features, and the outputs from the two groups are pooled together to generate the final output.

\begin{table}[]
    \centering
    \caption{Impact of audio-visual feature fusion on classification performance}
    \label{tab:ModDepthvsmAP}
    \begin{tabular}{cc}
        \toprule 
        Modality specific depth & mAP \\ \midrule
        2 & 49.6\\
        3 & 50.6\\
        4 & 50.0\\ \bottomrule
    \end{tabular}
    \vspace{-3mm}
\end{table}

As observed in Table~\ref{tab:ModDepthvsmAP}, both very shallow and very deep fusion result in lower performance. Note that the majority of the parameters are in the third and fourth layers as the feature dimension increases across groups. Thus, fusion after the second group means that the majority of the parameters are used for joint modeling, while a small fraction is used for modality-specific modeling. This implies that the audio-visual models need some modality-specific layers to map the audio and visual embeddings into a common space, which the joint layers can utilize. In the future, we want to carry out this experiment at a much finer level, i.e., layer by layer, as opposed to group by group. In the subsequent experiments, we fuse the audio and visual features after the third group. We no longer force both the audio and visual encoders to have four groups, and the final architecture is shown in Figure~\ref{fig:DAVSS_arch}.

\subsection{Knowledge distillation and initialization}
In this section, we explore the impact of model initialization and knowledge distillation on the DAVSS performance. As seen in Table~\ref{tab:pretrain_KD}, knowledge distillation significantly improves the performance, but the ImageNet initialization is needed to achieve the best results. These observations are consistent with distilled audio state-space models. 

Here, we perform an additional experiment to determine whether initializing the audio branch in the DAVSS model further improves the performance. Based on the experimental results, initializing the audio encoder with an audio-pretrained model improves the performance. This also increases the convergence rate of the model training. 

Previous work on distilled audio state-space models has observed that using a combination of teachers or using a stronger teacher model improves the performance of the student model~\cite{bhati2024dass}. We want to further explore if initializing the audio encoder with an even stronger variant of the DASS model improves the performance. We use  DASSv3, DASS model trained with SSLAM as the teacher model which outperforms base DASS model. Experiments show that the performance saturates for DASS initialization with the two initialization achieving similar performance. 

\vspace{-3mm}
\begin{table}[h!]
    \centering
    \caption{Impact of Knowledge distillation (KD) and pretrained the audio and visual encoders on performance on AudioSet-2 M.}
    \label{tab:pretrain_KD}
    \begin{tabular}{cccc}
    \toprule
        Audio init & Image init & KD & mAP \\ \midrule
        random & random & False & 28.2\\
        random & random & True & 47.6\\
        Imagenet & ImageNet & True & 50.3\\
        DASS & Imagenet & True & 52.2\\
        DASSv3 & Imagenet & True & 52.1\\ \bottomrule
    \end{tabular}
\end{table}
\vspace{-3mm}

\subsection{Stronger teachers: Multi-frame CAV-MAE}

Audio experiments have shown that using stronger audio leads to stronger, better-performing students~\cite{bhati2024dass}. We want to explore if the same trend is true for audio-visual models. 

To the best of our knowledge, CAV-MAE is the strongest open-source model available. To make a stronger CAV-MAE teacher, we utilize the observation that merging the predictions from multiple frames increases performance over using a single frame. Currently, a single frame is used as input by the CAV-MAE teacher for distillation, and we want to increase the number of frames while distilling the model. While an increase in the number of frames leads to increased performance, the performance saturates after using 4-5 frames~\cite{gongcontrastive}. We uniformly sampled 5 frames from the 10 total frames.  

\begin{table}[]
    \centering
    \caption{Impact of using stronger teachers on student performance}
    \label{tab:strong_teach}
    \begin{tabular}{ccc} \toprule
        Teacher & frame-ensemble & mAP \\ \midrule
        Single-frame & No & 51.5 \\
        Single-frame & Yes & 52.2 \\
        Multiple-frames(DASSv3) & No & 52.2 \\
        Multiple-frames(DASSv3) & Yes & 52.5 \\
        \bottomrule
    \end{tabular}
    \vspace{-2mm}
\end{table}

As seen in Table~\ref{tab:strong_teach}, using multiple frames increases the performance of the student DAVSS model when a single frame is used for inference.  However, for the typical inference setting, i.e., frame-ensemble over 10 frames, the performance of the two models is almost similar. In the future, we would like to explore using stronger audio-visual models outside of the CAV-MAE model family. 

\subsection{Understanding the performance gains}
DAVSS models, while being smaller, still outperform the larger transformer-based models. We hypothesize that the DAVSS models achieve high performance because they process the input audio-visual data with a small patch size or at a very fine resolution. 
The transformer-based models use a coarser resolution, i.e., larger patch size, to reduce the resulting input sequence lengths. To test this hypothesis, we train DAVSS models with various resolutions. The audio-visual classification performance decreases as the resolution of the DAVSS models becomes coarser, as observed in Figure~\ref{fig:DASS_overview}(b). These experimental results support our hypothesis that the performance gains of the state-space models are due to operating at a finer resolution compared to transformer-based alternatives. 

\subsection{Comparison with other methods}

In this section, we compare the DAVSS model with existing methods on the audio-visual classification task.   Table~\ref{tab:main_mAP} shows that DAVSS models, while being significantly smaller than the existing transformer-based models, outperform much larger models such as CAV-MAE and Audiovisual MAE. DAVSS models distilled from the CAV-MAE model even outperform the teacher models on both AudioSet and VGGSound dataset. 
DAVSS-tiny, a model with 12 times less parameters achieves comparable performance to most transformers based methods and even outperforms CAV-MAE on both AudioSet and VGGSound benchmarks. DAVSS models combine the performance of transformer models with the efficiency of SSMs. 

\begin{table}[h!]
    \centering
    \caption{Audio-visual event classification performance on AS(AudioSet) and VS(VGGSound) datasets. $*$ Approximation, both audio and visual encoders are ViT-B (87 Million parameters)}
    \label{tab:main_mAP}
    \begin{tabular}{cccc} \toprule
         & Params & AS (mAP) & VS(acc) \\ 
         CAV-MAE~\cite{gongcontrastive} & 165 & 50.5 & 65.4 \\ 
         CAV-MAE$^{scale+}$~\cite{gongcontrastive} & 165 & 51.2 & 65.5\\
         Audiovisual MAE~\cite{georgescu2023audiovisual} & 170$^{*}$  & 51.8 & 65.0\\
         MAViL~\cite{huang2023mavil} & 170$^{*}$  & 53.3 & 67.1\\
         EquiAV~\cite{kim2024equiav} & 173  & 54.6 & 67.1\\
         DAVSS-Tiny & 14 & 51.6 & 65.4\\
         DAVSS-Small & 47 &  52.5 & 66.7\\ 
         DAVSS-Medium & 84 & 52.7 & 67.5 \\
         \bottomrule
    \end{tabular}
\vspace{-3mm}
\end{table}

\section{Conclusions and Future Work}

We present DAVSS, a state-space-based audio-visual acoustic event classification model. DAVSS is significantly smaller than the existing transformer-based models, such as CAV-MAE, and yet still outperforms the latter. DAVSS models make up for a smaller model size by increasing the sequence lengths or processing the data at a finer resolution compared to CAV-MAE models. Varying the resolution also enables us to control the trade-off between throughput and the accuracy of the model.

\section{Generative AI Use Disclosure}
Generative AI was only used for basic editing such as grammar checks and improvements. 

\bibliographystyle{IEEEtran}
\bibliography{refs}

@article{gu2021efficiently,
  title={Efficiently modeling long sequences with structured state spaces},
  author={Gu, Albert and Goel, Karan and R{\'e}, Christopher},
  journal={arXiv preprint arXiv:2111.00396},
  year={2021}
}

@article{gu2023mamba,
  title={Mamba: Linear-time sequence modeling with selective state spaces},
  author={Gu, Albert and Dao, Tri},
  journal={arXiv preprint arXiv:2312.00752},
  year={2023}
}

@article{hamza2024audio,
  title={Audio Mamba: Bidirectional State Space Model for Audio Representation Learning},
  author={Hamza Erol, Mehmet and Senocak, Arda and Feng, Jiu and Son Chung, Joon},
  journal={arXiv e-prints},
  pages={arXiv--2406},
  year={2024}
}

@article{lin2024audio,
  title={Audio Mamba: Pretrained Audio State Space Model For Audio Tagging},
  author={Lin, Jiaju and Hu, Haoxuan},
  journal={arXiv preprint arXiv:2405.13636},
  year={2024}
}

@article{zhu2024vision,
  title={Vision mamba: Efficient visual representation learning with bidirectional state space model},
  author={Zhu, Lianghui and Liao, Bencheng and Zhang, Qian and Wang, Xinlong and Liu, Wenyu and Wang, Xinggang},
  journal={arXiv preprint arXiv:2401.09417},
  year={2024}
}

@article{liu2024vmamba,
  title={Vmamba: Visual state space model},
  author={Liu, Yue and Tian, Yunjie and Zhao, Yuzhong and Yu, Hongtian and Xie, Lingxi and Wang, Yaowei and Ye, Qixiang and Liu, Yunfan},
  journal={arXiv preprint arXiv:2401.10166},
  year={2024}
}

@inproceedings{gemmeke2017audio,
  title={Audio set: An ontology and human-labeled dataset for audio events},
  author={Gemmeke, Jort F and Ellis, Daniel PW and Freedman, Dylan and Jansen, Aren and Lawrence, Wade and Moore, R Channing and Plakal, Manoj and Ritter, Marvin},
  booktitle={2017 IEEE international conference on acoustics, speech and signal processing (ICASSP)},
  pages={776--780},
  year={2017},
  organization={IEEE}
}

@article{huang2023mavil,
  title={Mavil: Masked audio-video learners},
  author={Huang, Po-Yao and Sharma, Vasu and Xu, Hu and Ryali, Chaitanya and Li, Yanghao and Li, Shang-Wen and Ghosh, Gargi and Malik, Jitendra and Feichtenhofer, Christoph and others},
  journal={Advances in Neural Information Processing Systems},
  volume={36},
  pages={20371--20393},
  year={2023}
}

@inproceedings{gongcontrastive,
  title={Contrastive Audio-Visual Masked Autoencoder},
  author={Gong, Yuan and Rouditchenko, Andrew and Liu, Alexander H and Harwath, David and Karlinsky, Leonid and Kuehne, Hilde and Glass, James R},
  booktitle={The Eleventh International Conference on Learning Representations}
}

@inproceedings{georgescu2023audiovisual,
  title={Audiovisual masked autoencoders},
  author={Georgescu, Mariana-Iuliana and Fonseca, Eduardo and Ionescu, Radu Tudor and Lucic, Mario and Schmid, Cordelia and Arnab, Anurag},
  booktitle={Proceedings of the IEEE/CVF International Conference on Computer Vision},
  pages={16144--16154},
  year={2023}
}

@article{haliassos2022jointly,
  title={Jointly learning visual and auditory speech representations from raw data},
  author={Haliassos, Alexandros and Ma, Pingchuan and Mira, Rodrigo and Petridis, Stavros and Pantic, Maja},
  journal={arXiv preprint arXiv:2212.06246},
  year={2022}
}

@article{bhati2024dass,
  title={DASS: Distilled Audio State Space Models Are Stronger and More Duration-Scalable Learners},
  author={Bhati, Saurabhchand and Gong, Yuan and Karlinsky, Leonid and Kuehne, Hilde and Feris, Rogerio and Glass, James},
  journal={arXiv preprint arXiv:2407.04082},
  year={2024}
}

@inproceedings{jiang2025speech,
  title={Speech slytherin: Examining the performance and efficiency of mamba for speech separation, recognition, and synthesis},
  author={Jiang, Xilin and Li, Yinghao Aaron and Florea, Adrian Nicolas and Han, Cong and Mesgarani, Nima},
  booktitle={ICASSP 2025-2025 IEEE International Conference on Acoustics, Speech and Signal Processing (ICASSP)},
  pages={1--5},
  year={2025},
  organization={IEEE}
}

@article{zhang2025mamba,
  title={Mamba in speech: Towards an alternative to self-attention},
  author={Zhang, Xiangyu and Zhang, Qiquan and Liu, Hexin and Xiao, Tianyi and Qian, Xinyuan and Ahmed, Beena and Ambikairajah, Eliathamby and Li, Haizhou and Epps, Julien},
  journal={IEEE Transactions on Audio, Speech and Language Processing},
  year={2025},
  publisher={IEEE}
}

@inproceedings{sarkar2023self,
  title={Self-supervised audio-visual representation learning with relaxed cross-modal synchronicity},
  author={Sarkar, Pritam and Etemad, Ali},
  booktitle={Proceedings of the AAAI Conference on Artificial Intelligence},
  volume={37},
  number={8},
  pages={9723--9732},
  year={2023}
}

@inproceedings{morgado2021audio,
  title={Audio-visual instance discrimination with cross-modal agreement},
  author={Morgado, Pedro and Vasconcelos, Nuno and Misra, Ishan},
  booktitle={Proceedings of the IEEE/CVF conference on computer vision and pattern recognition},
  pages={12475--12486},
  year={2021}
}

@article{kim2024equiav,
  title={EquiAV: leveraging equivariance for audio-visual contrastive learning},
  author={Kim, Jongsuk and Lee, Hyeongkeun and Rho, Kyeongha and Kim, Junmo and Chung, Joon Son},
  journal={arXiv preprint arXiv:2403.09502},
  year={2024}
}

@inproceedings{guo2024crossmae,
  title={Crossmae: Cross-modality masked autoencoders for region-aware audio-visual pre-training},
  author={Guo, Yuxin and Sun, Siyang and Ma, Shuailei and Zheng, Kecheng and Bao, Xiaoyi and Ma, Shijie and Zou, Wei and Zheng, Yun},
  booktitle={Proceedings of the IEEE/CVF Conference on Computer Vision and Pattern Recognition},
  pages={26721--26731},
  year={2024}
}

@article{araujo2025cav,
  title={CAV-MAE Sync: Improving Contrastive Audio-Visual Mask Autoencoders via Fine-Grained Alignment},
  author={Araujo, Edson and Rouditchenko, Andrew and Gong, Yuan and Bhati, Saurabhchand and Thomas, Samuel and Kingsbury, Brian and Karlinsky, Leonid and Feris, Rogerio and Glass, James R},
  journal={arXiv preprint arXiv:2505.01237},
  year={2025}
}

@article{xing2024locality,
  title={Locality-aware Cross-modal Correspondence Learning for Dense Audio-Visual Events Localization},
  author={Xing, Ling and Qu, Hongyu and Yan, Rui and Shu, Xiangbo and Tang, Jinhui},
  journal={arXiv preprint arXiv:2409.07967},
  year={2024}
}

@article{aytar2016soundnet,
  title={Soundnet: Learning sound representations from unlabeled video},
  author={Aytar, Yusuf and Vondrick, Carl and Torralba, Antonio},
  journal={Advances in neural information processing systems},
  volume={29},
  year={2016}
}

@article{dao2024transformers,
  title={Transformers are ssms: Generalized models and efficient algorithms through structured state space duality},
  author={Dao, Tri and Gu, Albert},
  journal={arXiv preprint arXiv:2405.21060},
  year={2024}
}

@inproceedings{lin2024siamese,
  title={Siamese vision transformers are scalable audio-visual learners},
  author={Lin, Yan-Bo and Bertasius, Gedas},
  booktitle={European Conference on Computer Vision},
  pages={303--321},
  year={2024},
  organization={Springer}
}

@inproceedings{afouras2020self,
  title={Self-supervised learning of audio-visual objects from video},
  author={Afouras, Triantafyllos and Owens, Andrew and Chung, Joon Son and Zisserman, Andrew},
  booktitle={Computer Vision--ECCV 2020: 16th European Conference, Glasgow, UK, August 23--28, 2020, Proceedings, Part XVIII 16},
  pages={208--224},
  year={2020},
  organization={Springer}
}

@inproceedings{arandjelovic2017look,
  title={Look, listen and learn},
  author={Arandjelovic, Relja and Zisserman, Andrew},
  booktitle={Proceedings of the IEEE international conference on computer vision},
  pages={609--617},
  year={2017}
}

@inproceedings{chen2020vggsound,
  title={Vggsound: A large-scale audio-visual dataset},
  author={Chen, Honglie and Xie, Weidi and Vedaldi, Andrea and Zisserman, Andrew},
  booktitle={ICASSP 2020-2020 IEEE International Conference on Acoustics, Speech and Signal Processing (ICASSP)},
  pages={721--725},
  year={2020},
  organization={IEEE}
}

\end{document}